\documentclass[acus]{JAC2000}

\usepackage{graphicx}

\begin{document}
\title{
LARGE-SCALE SIMULATION OF BEAM DYNAMICS IN HIGH INTENSITY ION
LINACS USING PARALLEL SUPERCOMPUTERS\thanks{Work supported by the
DOE Grand Challenge in Computational Accelerator Physics,
Advanced Computing for 21st Century Accelerator Science and Technology
Project, and the Los Alamos Accelerator Code Group using resources at
the Advanced Computing Laboratory and the National Energy
Research Scientific Computing Center.}}

\author{Robert D. Ryne and Ji Qiang, LANL, Los Alamos, NM 87545, USA}

\maketitle

\begin{abstract} 
In this paper we present results of using parallel supercomputers
to simulate beam dynamics in next-generation high intensity ion linacs.
Our approach uses a
three-dimensional space charge calculation with six types of boundary
conditions.
The simulations use a hybrid approach involving transfer maps to treat
externally applied fields
(including rf cavities)
and parallel particle-in-cell techniques to
treat the space-charge fields. 
The large-scale simulation results presented here represent a three
order of magnitude improvement in simulation capability, in terms of
problem size and speed of execution, compared with typical two-dimensional
serial simulations.
Specific examples
will be presented, including simulation of the spallation neutron source
(SNS) linac and the Low Energy Demonstrator Accelerator (LEDA) beam halo
experiment.
\end{abstract}

\section{INTRODUCTION}
The high intensity of future accelerator-driven systems places
stringent requirements on the allowed beam loss, since very
small fractional losses at high energy can produce unacceptably high
levels of radioactivity.
Previous studies suggest that the low density, large amplitude halo of
the beam is a major
issue for these systems~\cite{gluckstern1,wangler,qiang0}.
Large-scale simulations are an important tool for exploring the
beam dynamics, predicting the beam halo, and facilitating design decisions
aimed at controlling particle loss and meeting operational requirements.

The most widely used model for simulating intense beams in ion rf linacs
is represented by the Poisson-Vlasov equations.
These equations are often solved using a particle-in-cell (PIC) approach.
In this paper we will describe a parallel simulation capability
that combines the PIC method with techniques from magnetic optics,
and we will present results of using parallel supercomputers
to simulate beam dynamics in high intensity ion rf linacs.

%
\section{PHYSICAL MODEL AND NUMERICAL METHODS}
In the
PIC
approach a number of simulation particles, 
called macroparticles, are used to solve (indirectly) the evolution 
equations and model the charged particle dynamics. The motion of
individual particles in the absence of radiation can be described by
Hamilton's equations,
\begin{equation}
{d{\vec q}\over dt}= {\partial H \over \partial {\vec
p}},~~~~~~~~~~
{d{\vec p}\over dt}= -{\partial H \over \partial {\vec q}},
\end{equation}
where $H({\vec q},{\vec p},t)$ denotes the Hamiltonian of the system,
and where ${\vec q}$ and ${\vec p}$ denote canonical coordinates and
momenta, respectively.
In the language of {\em mappings}
we would say that there is a
(generally nonlinear) map, ${\cal M}$, corresponding to the
Hamiltonian $H$, which maps initial phase space variables, $\zeta^i$,
into final variables, $\zeta^f$, and we write
\begin{equation}
\zeta^f={\cal M}\zeta^i.
\end{equation}
The potential in the Hamiltonian includes contributions from both
the external fields and the space-charge fields.
In the Poisson-Vlasov approach, discreteness effects are neglected
and the space charge is represented by a smoothly varying mean field.
Typically, the Hamiltonian 
can be written as a sum of two parts, $H = H_{ext} + H_{sc}$,
which correspond to the external and space-charge contributions.
Such a situation is ideally suited to multi-map
symplectic split-operator methods~\cite{forestall}.
A second-order-accurate algorithm for a single step is given by
\begin{equation}
{\cal M}(\tau)={\cal M}_1(\tau/2)~{\cal M}_2(\tau)~{\cal M}_1(\tau/2)~,
\label{splop2}
\end{equation}
where $\tau$ denotes the step size, ${\cal M}_1$ is the map corresponding
to $H_{ext}$ and ${\cal M}_2$ is the map corresponding to $H_{sc}$.
This approach can be
easily generalized to higher order accuracy using Yoshida's
scheme if desired~\cite{yoshida}.

The electrostatic scalar potential generated by
the charged particles is obtained by solving Poisson's
equation
\begin{equation}
\nabla^2 \Psi({\bf r}) =-\rho({\bf r})/\epsilon_0.
\label{poisson}
\end{equation}
where $\rho$ is the charge density.
We have developed a Fourier-based transformation and
an eigenfunction expansion method to handle six
different boundary conditions:
(1) open in all three dimensions; (2) open transversely and periodic
longitudinally; (3,4) round conducting pipe transversely and
open or periodic longitudinally;  (5,6) rectangular conducting pipe
transversely and open or periodic longitudinally.
A discussion of the numerical algorithms for solving the Poisson's equation
with these different boundary conditions can be found in~\cite{qiang}.

The charge density $\rho$ on the grid
is obtained by using a volume-weighted linear interpolation
scheme\cite{hockney,birdsall}.
After the potential and electric field is found on the grid, the same
scheme is used to interpolate the field at the particle locations. 
During the course of the simulation each step involves the following:
transport of a numerical distribution of particles through a half step
based on ${\cal M}_1$, 
solving Poisson's
equation based on the particle positions and performing a space-charge
``kick'' ${\cal M}_2$,
and performing transport
through the remaining half of the step based on ${\cal M}_1$.

\section{APPLICATIONS}
We have applied the above 3D parallel PIC approach to
an early design of the SNS linac and to
the proposed LEDA beam halo experiment.
Our simulation of the SNS linac starts at the beginning of the
DTL.
The code advances particles through drift spaces, quadrupole fields
and RF gaps. The dynamics inside the gaps is computed using external
fields calculated from the electromagnetic code SUPERFISH~\cite{superfish}.
A schematic plot of the SNS linac configuration used in this study
is shown in Figure~1~\cite{bhatia}.
It consists of three types of RF structures:
a DTL, a CCDTL, and a CCL.
There are a total of
425 RF segments in the linac. Figure 2 shows the rms transverse size
($x_{rms},y_{rms}$) and the maximum transverse extent
($x_{max},y_{max}$) of the bunched beam in the linac with one set of errors.
We see that the maximum particle amplitude is well-below the
aperture size of the linac. This margin is needed to operate the
linac safely and to avoid beam loss at the high energy end.
The jump in rms beam size between
the DTL and CCDTL at $20$ MeV is due to a change of
focusing period from $8\beta \lambda$ to $12 \beta \lambda$ at
$805$ MHz.
\vspace{-.05in}
\begin{figure}[htb]
\centering
\includegraphics*[width=80mm,angle=0]{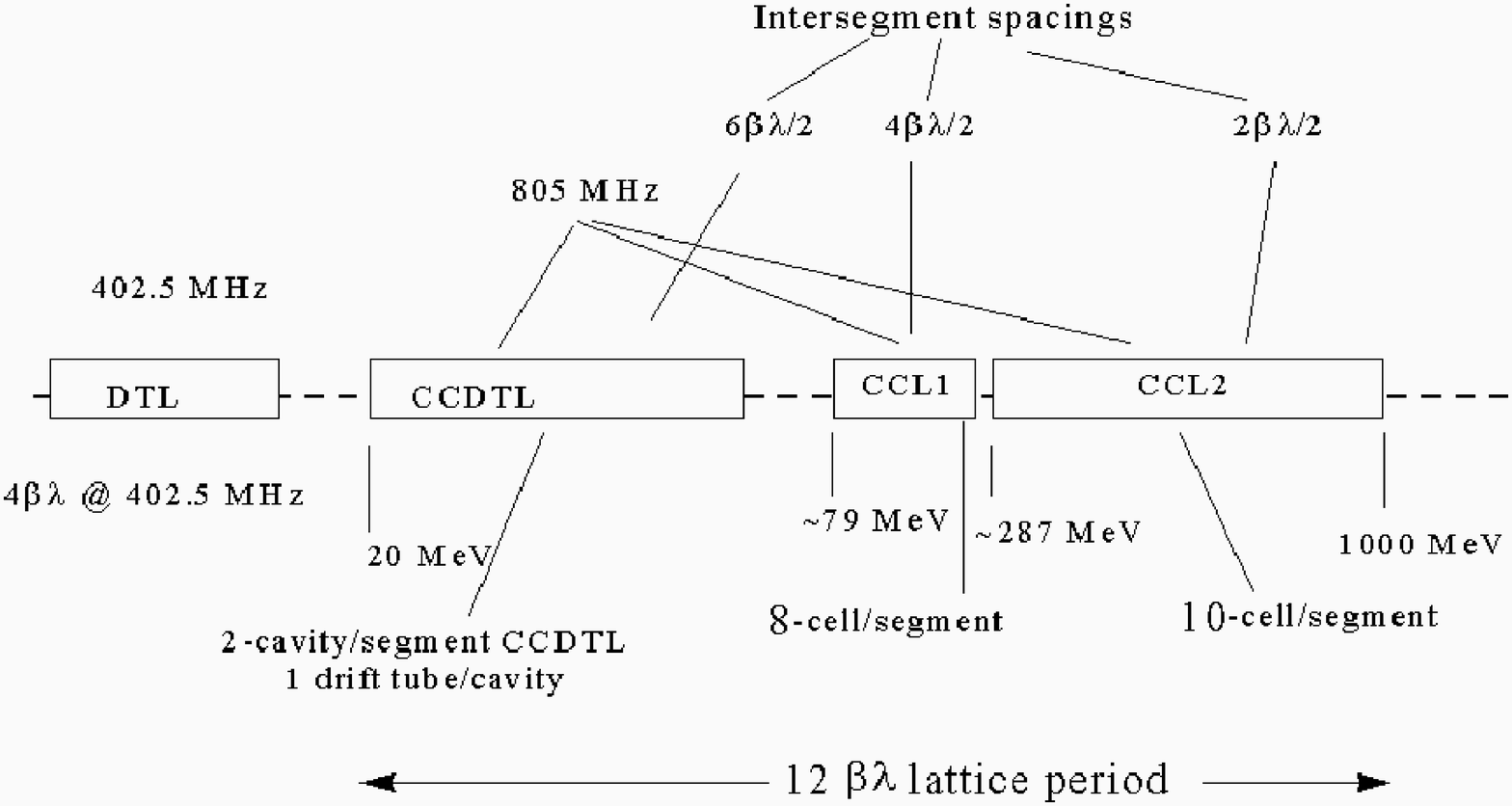}
\caption{The SNS linac configuration}
\end{figure}
\begin{figure}[htb]
\centering
\includegraphics*[width=55mm,angle=-90]{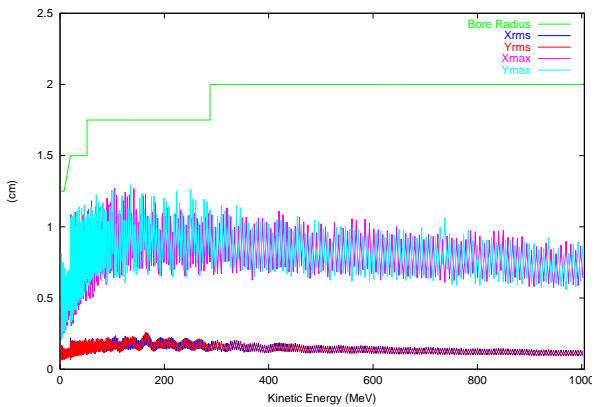}
\caption{Transverse beam size as a function of kinetic energy in the SNS 
linac}
\end{figure}

~\
In the LEDA beam halo experiment, a mismatched high-intensity proton beam
will be propagated through a periodic focusing transport system
and measurements will be made of the beam profile.
The goals of the experiment are two-fold: first, to study beam halo
formation and test our physical understanding of the phenomena, and
second, to evaluate our computational models and assess their
predictive capability through a comparison of simulation and experiment.
Fig.~3 gives a schematic plot of the layout of the experiment~\cite{wangler2}.
It consists of 52 alternating-focusing quadrupole magnets with a focusing
period of $41.96$ cm. The gradients of the first four quadrupole
magnets can be adjusted to create a mismatch that excites the breathing
mode or the quadrupole mode. The transverse beam profile will be
measured using a beam-profile scanner.
Fig.~4 and Fig.~5 present simulation results of the transverse
beam size for the breathing mode and the quadrupole mode,
plotted at the center of the drift spaces between quadrupole magnets,
as a function of distance. The plots include both the rms beam size
and the maximum particle extent in the simulation.
The physical parameters for the simulation
were I=$100$ mA, E=6$.7$ MeV, and f=$350$ MHz.
The simulation was performed using $100$ million macroparticles
with a 128x128x256 (x-y-z) space-charge grid.
\vspace{-.1in}
\begin{figure}[htb]
\centering
\includegraphics*[width=80mm,angle=0]{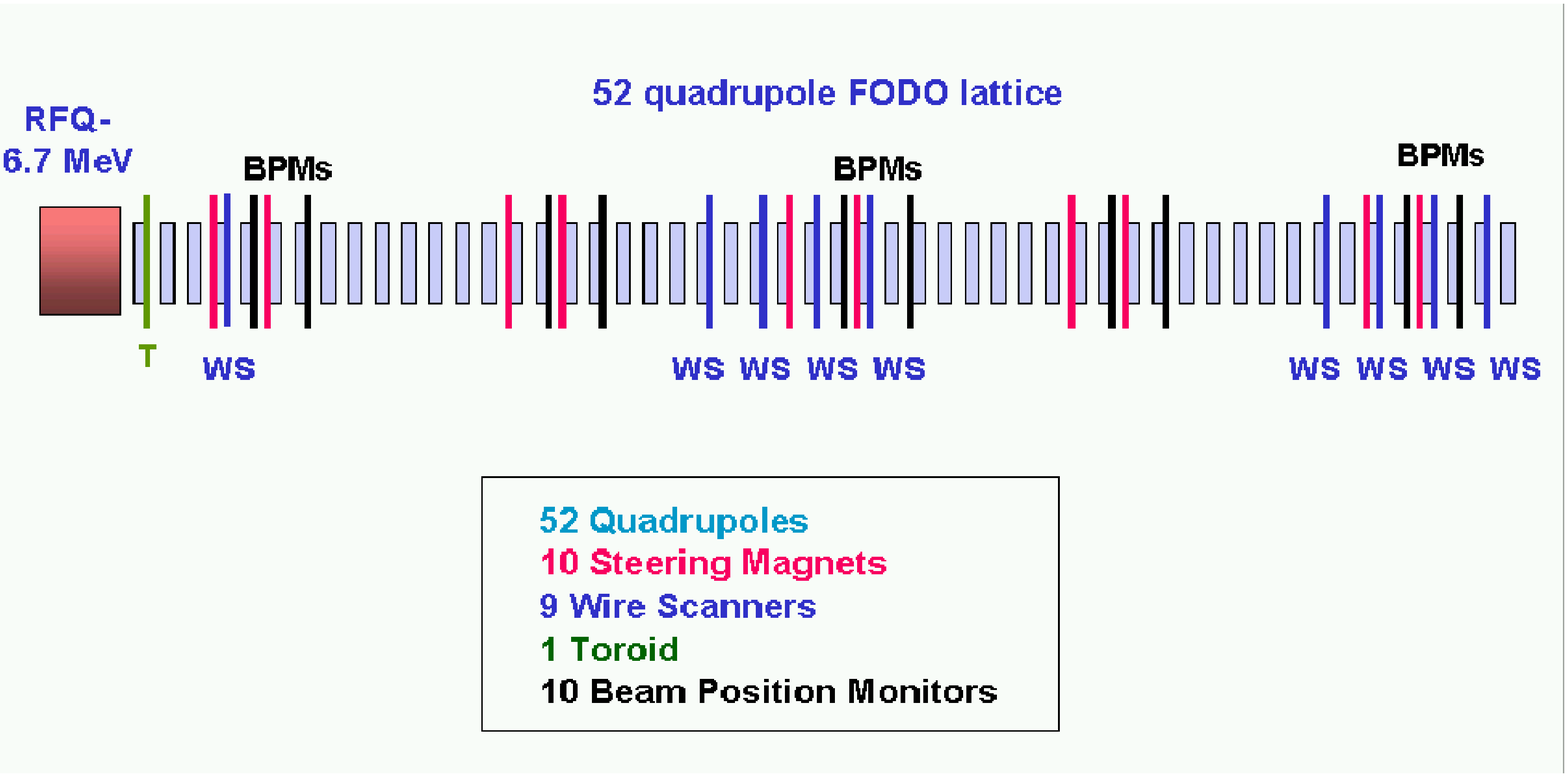}
\caption{LEDA halo experiment layout}
\end{figure}
\begin{figure}[htb]
\centering
\includegraphics*[width=55mm,angle=-90]{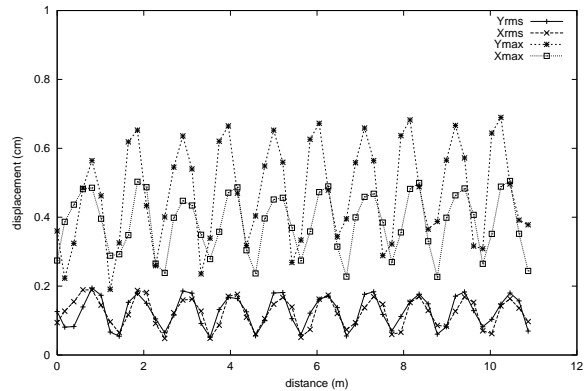}
\caption{
Transverse beam size as a function of
distance for the breathing mode in the LEDA halo experiment}
\end{figure}
\begin{figure}[htb]
\centering
\includegraphics*[width=55mm,angle=-90]{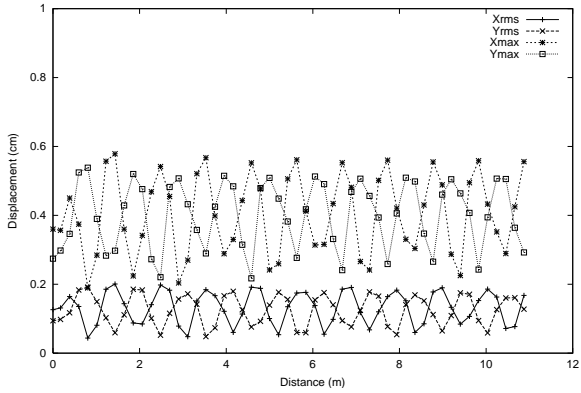}
\caption{
Transverse beam size as a function of
distance for the quadrupole mode in the LEDA halo experiment}
\end{figure}

From Fig.~4, the two transverse components of the breathing mode
are in phase, while the quadrupole mode in Fig.~5 has the two
components out of phase. Evidently, it will be possible
in the experiment to clearly excite either of the two modes.
Furthermore, the debunching of the beam will not
significanly alter the structure of the oscillations.
Fig.~6 shows the accumulated one-dimensional density profiles (along $x$)
for the breathing mode and the quadrupole mode just after the magnet \#$49$. 
\begin{figure}
\centering
\includegraphics*[width=55mm,angle=-90]{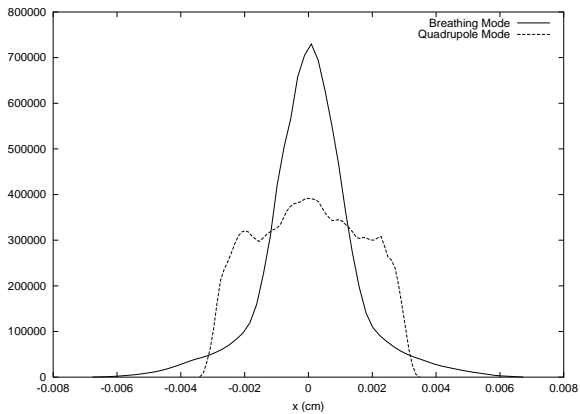}
\caption{
Accumulated density profile along $x$ for the breathing mode and the
quadrupole mode}
\end{figure}
The breathing mode is more peaked and has a larger extent than the
quadrupole mode.
Measurements will be taken at this location and
will be compared with our simulations. 
The data in Fig.~6 are well-resolved over a range of about $6$ decades.

An important piece of information from a design standpoint is
amount of charge beyond a specified radius or spatial location
as a function of distance along the accelerator.
This is shown graphically in Fig.~7 which shows the complement of
the horizontal and vertical cumulative density profiles,
at every step, when the quadrupole mode is excited.
In other words, the contours describe the fraction of charge that
would be intercepted by a scraper placed at that transverse position.
\begin{figure}[htb]
\centering
\includegraphics*[width=80mm,angle=-90]{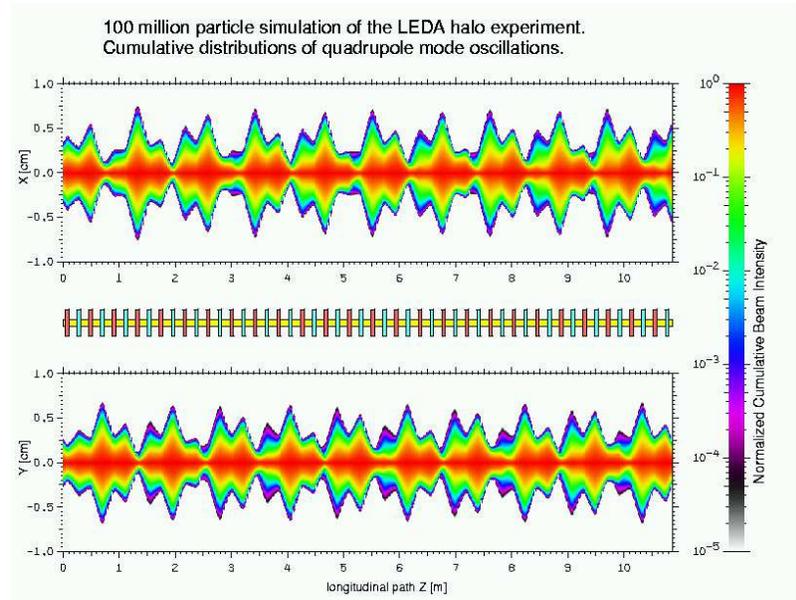}
\caption{Horizontal and vertical cumulative density profiles
of a quadrupole mode mismatch in the LEDA halo experiment.}
\end{figure}

The above LEDA simulations
used 100 million macroparticles and a 3D Poisson solver,
and required only 2 hours to execute on 256 processors.
In contrast, beam dynamics simulations performed on serial
computers typically use 10,000 to 100,000 macroparticles and a
2D Poisson solver. Even if the above large-scale calculations could be
performed on a PC, they would require on the order of a month to complete.
In conclusion, while small-scale simulations on serial computers
are extremely valuable for rapid design and predicting rms properties,
large-scale simulations are needed for high-resolution studies
aimed at making quantitative predictions of the beam halo.

\section{Acknowledgments}
We thank the SNS linac design team and the
LEDA beam halo experiment team for helpful discussions.
We also thank S. Habib for helpful discussions and
C. Thomas Mottershead for graphics support.

\end{document}